\documentclass[a4paper]{article}
\usepackage[utf8]{inputenc}
\usepackage{amsmath}
\usepackage{amsfonts}
\usepackage{amssymb}
\usepackage{graphicx}
\usepackage{multirow}
\usepackage{geometry}
\usepackage{subcaption}
\usepackage{xcolor}

\numberwithin{theo}{section} 

\usepackage{url}

\begin{document}

\begin{center}
\section*{Modelling fungal hypha tip growth\\ via viscous sheet approximation}
\large{T. de Jong}

\small{Center for Analysis, Scientific Computing and Applications (CASA),\footnote{former affiliation}\\
TU  Eindhoven,
P.O Box 513, Eindhoven, The Netherlands\\
{\tt t.g.de.jong.math@gmail.com}}
\\[2mm]

\large{J. Hulshof}

\small{Department of Mathematics, Faculty of Sciences\\
VU University Amsterdam, De Boelelaan 1105, 1081 HV Amsterdam\\
{\tt j.hulshof@vu.nl}}\\[2mm]

\large{G. Prokert\footnote{corresponding author}}

\small{Center for Analysis, Scientific Computing and Applications (CASA),\\
TU  Eindhoven, P.O Box 513, Eindhoven, The Netherlands\\
{\tt g.prokert@tue.nl}}\\[2mm]

\end{center}

%
%

\begin{small} \textbf{Abstract.} In this paper we present a new model for single-celled, non-branching hypha tip growth. The growth mechanism of hypha cells consists of transport of cell wall building material to the cell wall and subsequent incorporation of this material in the wall as it arrives.  To model the transport of cell wall building material to the cell wall we  follow Bartnicki-Garcia et al in assuming that the cell wall building material is transported in straight lines by an isotropic point source. To model the dynamics of the cell wall, including its growth by new material,  we use the approach of Campas and Mahadevan, which assumes that the cell wall is a thin viscous sheet sustained by a pressure difference. Furthermore, we include a novel equation which models the hardening of the cell wall as it ages.  We present  numerical results which give evidence that our model can describe tip growth, and briefly discuss validation aspects.

\vspace{3mm}
{\bf MSC:} 92C10, 76Z99

{\bf Keywords:} hypha growth, viscous sheet, cell wall growth
\end{small}
%
%

\section{Introduction} 

\begin{figure}[h]
\begin{center}
\includegraphics[width=13cm]{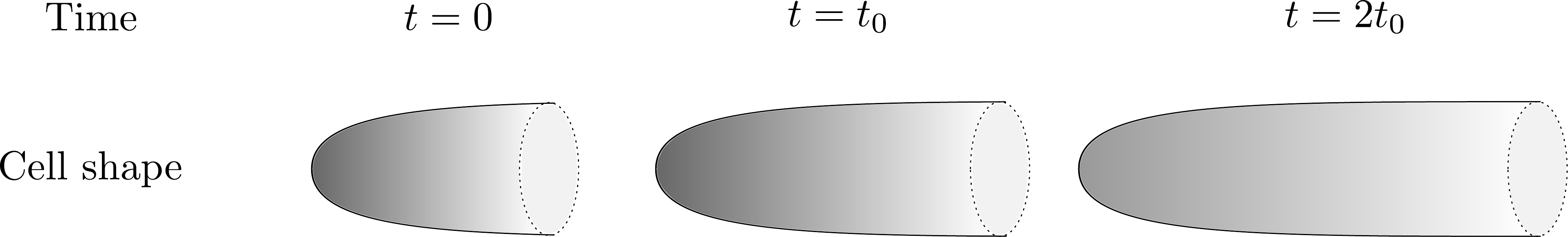}
\caption{Idealized hyphae growth: The figures display the growth of the cell wall at time steps $t_0$.  Qualitatively the shape remains unchanged during tip growth.  \label{fig:hypha}}\end{center} \end{figure}

In this article we model  the growth of fungal filaments. Fungi filaments are called hyphae. Hyphae grow by localized cell extension at their tips.  During tip growth a hypha cell exhibits extreme lengthwise growth while its shape remains qualitatively the same and the tip's velocity remains approximately constant. Furthermore, in the absence of spatial influences the cell's shape is almost rotationally symmetric. In Figure \ref{fig:hypha} we display an idealized cell wall shape during tip growth.  \\

\begin{figure}[h]
\begin{center}
\includegraphics[width=7cm]{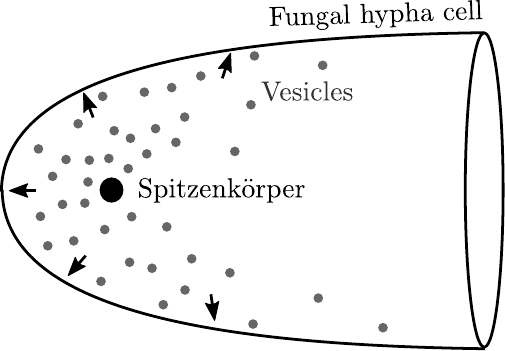}
\caption{A fungal cell with a Spitzenk\"{o}rper: Vesicles move from the Spitzenk\"{o}rper to the cell wall. When they hit the cell wall they are absorbed and the cell wall grows.  \label{fig:spitzie2}}\end{center} \end{figure}

In many fungi exhibiting tip growth a dynamic cluster of vesicles is present close to the tip \cite{GI57,MCC68,STE07}, see Figure \ref{fig:spitzie2}. It is called the Spitzenk\"{o}rper. It remains at an approximately fixed distance from the tip. The exact workings of the Spitzenk\"{o}rper are not understood. However, there is strong evidence that the Spitzenk\"{o}rper plays a crucial role in tip growth \cite{GI55,GI57,GI69}. These vesicles are transported to the cell wall. Fusion of the vesicles with the cell membrane leads to growth of the cell. Since the Spitzenk\"{o}rper is located close to the tip we expect most of the growth to take place at the tip.

Modelling of fungal tip growth  consists of two parts: transport of cell wall building material to the cell wall from the Spitzenk\"{o}rper and growth of the cell wall as new cell wall building material arrives.  Certain fungi follow approximately orthogonal growth trajectories \cite{BG00}. Assuming  that the growth is orthogonal and that the cell wall building material is transported in straight lines by an isotropic point source, Bartnicki-Garcia et al. arrived at a model which has the characteristic features of tip growth \cite{BG89}. Alternatively,  the transport of cell wall building materials can be modelled using a diffusive process as proposed by Koch \cite{KOCH94}. These models do not take into account the material properties of the cell wall. Experimental results from Wessels et al. \cite{WES83} suggest that the tip of the cell deforms  more easily than the part behind it. Campas and Mahadevan modelled this by assuming that the cell wall is a thin viscous sheet which increases in viscosity as the distance to the tip is increased \cite{CM09}. Goriely et al.  assumed an elastic response of the cell wall with decreasing elasticity away from the tip \cite{GO17,GO10}. Eggen et al. assume that the expansion of the cell wall is regulated by an expansion propensity which depends on an ageing process given by a constant rate Poissonian decay process \cite{EGG11}. These three models yield the characteristic features of tip growth.  \\

In this paper, we propose a new model that combines and extends the work of Bartnicki-Garcia et al. \cite{BG89}  and the work of Campas and Mahadevan \cite{CM09}. We assume
that the cell wall building material is transported in straight lines from an isotropic point source, and we describe the cell wall as a thin viscous sheet. Furthermore, we include a novel equation which models the hardening of the cell wall with age. We call our model the Ballistic Ageing Thin viscous Sheet model (BATS) model. The governing equations of this model are given by a five dimensional first order ODE. This model requires an explicit dependency of the viscosity on the age, called the viscosity function.  \\

 This paper is structured as follows: In Section \ref{secbio} we  discuss some background on the biology of fungal tip growth. In Section \ref{secmod} we describe the BATS model in detail. In Section \ref{secgov} we derive the governing equations as a first order five dimensional ODE and define which solutions correspond to tip growth.  In Section \ref{sec:numerics} we present some numerical examples  of tip growth  as described by the BATS model.\\

The main features of the BATS model have been announced (without a full derivation or discussion of the biological background) in \cite{JONGCOMFOS}.  The first author intends to describe the numerical aspects in detail in a forthcoming paper since the numerical aspects are of independent interest.

\section{Hyphal morphogenesis: Tip growth in fungi \label{secbio}}

When unaffected by external forces, a single non-branching hypha cell exhibiting tip growth has a tubular shape. It expands in the direction corresponding to the outward normal at its tip while its width away from the apical region remains almost unchanged.  During growth the tip's velocity remains approximately constant and the overall shape remains qualitatively the same, see Figure \ref{fig:hypha}. \\

In this section we give a brief exposition of the biology of tip growth in fungal hyphae. For a more in-depth treatment we refer to   \cite{KEIJ09,MO08}. \\

\subsection{The Spitzenk\"{o}rper}
\begin{figure}[h]
\begin{center}
\includegraphics[width=14cm]{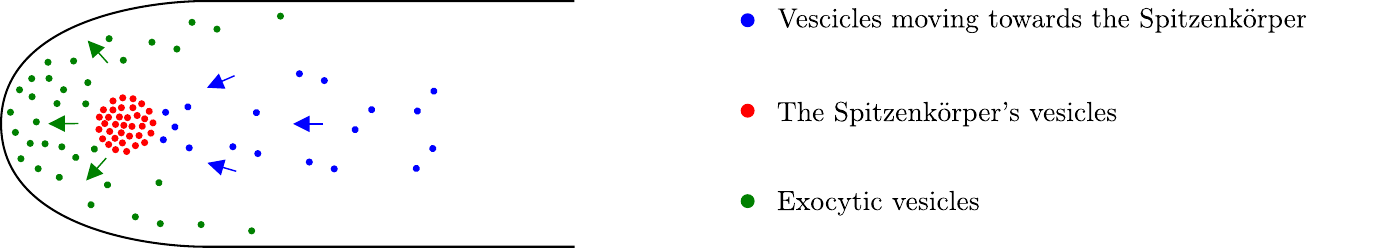}
\caption{Schematic of the hypothesized transport of vesicles over a fungal cell. We have omitted all hyphal cell contents besides vesicles. The newly produced vesicles get sent to the Spitzenk\"{o}rper. The Spitzenk\"{o}rper sends vesicles for exocytosis with the cell membrane. \label{fig:spitz}}\end{center} \end{figure}

In many fungi exhibiting tip growth a dynamic cluster of vesicles is present \cite{GI57,MCC68,STE07}. This cluster located close to the tip is called the Spitzenk\"{o}rper. It remains at an approximately fixed distance from the tip. The exact workings of the Spitzenk\"{o}rper are not understood. However, there is strong evidence that the Spitzenk\"{o}rper plays a crucial role in tip growth. Experiments by Girbardt with fungal hyphae of \textit{Polystictus} indicate that when the Spitzenk\"{o}rper  disappeared hyphal elongation stopped and when the Spitzenk\"{o}rper reappeared hyphal elongation continued \cite{GI55,GI57,GI69}. Furthermore, Girbardt observed that a change in the position of the Spitzenk\"{o}rper preceded a change in the growth direction of the hyphae \cite{GI69}. Hence, it has been hypothesized that the Spitzenk\"{o}rper controls hyphal elongation by sending vesicles to the cell wall \cite{BG89}. Experiments on the fungus \emph{Allomyces macrogynus} suggests that the vesicles at the Spitzenk\"{o}rper are synthesized far behind the hyphal tip and then collected by the Spitzenk\"{o}rper \cite{MCD00}. Vesicles destined for fusion with the cell membrane are called exocytic vesicles. Fusion with the cell membrane leads to growth of the cell wall and cell membrane.   The hypothesized vesicle transport is displayed in Figure \ref{fig:spitz}. \newline


%
%
%

In the fungus \textit{Neurospora crassa} the exocytic vesicles are tethered to the cell membrane along the so called exocyst complex. The exocyst complex seems to localize close to the hyphal tip \cite{RI14}. This gives strong evidence that due to the closeness of the Spitzenk\"{o}rper to the tip the apex receives most of the vesicles and consequently, grows the most. 

\subsection{Turgor pressure}
\begin{figure}[h]
\begin{center}
\includegraphics[width=8cm]{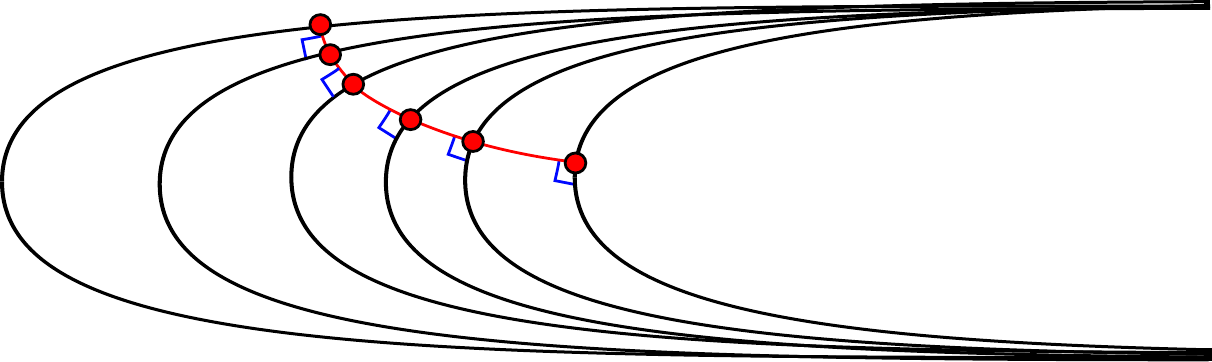}
\caption{Orthogonal growth trajectories.The black curves describe the growing cell wall. The trajectory that the red cell wall particle follows is represented by a red curve. This curve intersects the growing cell wall orthogonally. Hence, it is called an orthogonal growth trajectory.  \label{fig:ortho}}\end{center} \end{figure}

The cell wall thickness is of the same magnitude over the whole fungal cell. Consequently, the cell wall's area must expand, otherwise the cell wall would continuously thicken. In \cite{BG00} particles in the cell wall of the fungus \textit{Rhizoctonia solani} were tracked during tip growth. It turned out that the particles follow approximately orthogonal trajectories with respect to the cell wall, see Figure \ref{fig:ortho}. In addition, there is no indication that the cell rotates during tip growth.  

The observation that the cell wall particles follow orthogonal growth trajectories suggests that there is a driving force perpendicular to the cell wall surface. Under normal circumstances the hyphal contents exert a substantial pressure on the cell wall which yields a pressure difference between the inside and outside of the cell \cite{HUD92}. Puncturing the cell wall leads to  a rapid outflow of the cell's contents. Hence, the pressure difference might be the cause for orthogonal cell wall growth trajectories in \textit{Rhizoctonia solani}. However, no experimental results yield a direct link between tip growth and turgor pressure. Furthermore, there are biological arguments which suggest that the turgor pressure and the Spitzenk\"{o}rper are insufficient to explain tip growth:
\begin{itemize}
\item[-] \textbf{Oomycetes require no turgor pressure for tip growth:}  Oomycetes are water moulds belonging to the kingdom of Straminipila.  Tip growth in oomycetes can take place in the absence of measurable turgor pressure \cite{HA02,MOHA93}. However, for fungal hyphae it remains an unresolved issue whether tip growth can occur without turgor pressure  \cite{LEW07,LEW04}.
\item[-] \textbf{Orthogonal growth has only be shown for \textit{Rhizoctonia solani}:} The author is unaware whether the growth trajectories have been studied for any other fungal cell besides \textit{Rhizoctonia solani}. However, the root hairs of the plant \textit{Medicago truncatula} exhibit tip growth with non-orthogonal growth trajectories \cite{DU04}. 
\item[-] \textbf{Microscopic properties of the cell wall are important for the cell shape:} The microscopic properties of the cell wall determine how much stress the cell wall can bear and how it deforms.  Hence, these properties determine how the cell shape responds to a pressure difference.
\end{itemize}

The last point indicates that a description of the cell wall dynamics is necessary to model tip growth accurately. \\

\subsection{The cell wall dynamics: the soft spot}
To withstand the pressure difference the cell wall is generally very strong and rigid.  However, the chemical composition of the cell wall is not homogeneous.   In 1892 Reinhardt conjectured that cell wall close to the tip is more flexible compared to the part behind the tip. This is the so-called soft-spot hypothesis. Experimental evidence to support this claim was obtained more than 90 years later by Wessels et al. \cite{WES83}. Intuitively, this explains how the pressure difference allows for extreme growth in the longitudinal direction with hardly any expansion in the radial direction. Also observe that the flexibility/rigidity of the cell wall influences the growth trajectories. Consequently, non-orthogonal growth trajectories might be explained by the chemical composition of the cell wall.

\section{Derivation of the BATS model \label{secmod}}

In the previous section we presented biological evidence that the turgor pressure, the Spitzenk\"{o}rper and the cell wall dynamics drive tip growth. Hence, we have constructed a tip growth model which incorporates these biological components. We call our model the Ballistic Ageing Thin viscous Sheet (BATS) model. It models idealized tip growth since we assume there is an unlimited supply of cell wall building material. Hence, there is unlimited growth. We refer to this idealized tip growth as steady tip growth. In this section we will see that the BATS model consists of governing equations given by ODEs and an integral equation together with conditions which identify solutions corresponding to steady tip growth. We refer to these conditions as steady tip growth conditions. \\

%
The BATS model can be summarized as follows: 
\begin{enumerate}
\item \textbf{Steady tip growth is described by travelling wave profiles.} We assume that the cell is axially symmetric. We also assume that the cell grows at an approximately constant speed while preserving its overall shape. Hence, we model steady tip growth by an axially symmetric travelling wave profile. In addition, we formulate further conditions on the cell shape based on the biological cell shape of hypha cells.
\item \textbf{The cell wall is a thin viscous sheet sustained by the pressure difference.}  To model the evolution of the cell wall we use the work of Campas and Mahadevan \cite{CM09}. They assume that the cell wall is a thin viscous sheet sustained by the pressure difference between the inside and the outside of the cell. 
\item \textbf{The Spitzenk\"{o}rper is a ballistic vesicle supply center.} To model the Spitzenk\"{o}rper we use the work of Bartnicki-Garcia et al. \cite{BG89}. They assume  that the Spitzenk\"{o}rper is an isotropic point source which transports vesicles in straight lines at a constant rate to the cell wall.
\item \textbf{The cell wall viscosity increases with age.} 
We model the soft spot hypothesis by introducing a novel age variable. The age corresponds to the average age of all cell wall particles at a given location in the cell wall. We assume that the viscosity depends on the age. An explicit viscosity function needs to be chosen to complete the model. From the soft spot hypothesis we expect that the cell wall `hardens' with age. Hence, we consider viscosity functions which increase monotonously with age. 
The model of Campas and Mahadevan models the soft spot by assuming that the viscosity increases monotonously with the arclength to the tip \cite{CM09}. Eggen, Keijzer and Mulder have proposed an age equation in \cite{EGG11} but this differs from the age equation of the ballistic ageing thin viscous sheet model. 
\end{enumerate}

We devoted a subsection to each of these modelling components.



\subsection{Steady tip growth is described by travelling wave profiles \label{sec:stg}}

During tip growth the hypha's tip moves at an approximately constant speed and the hypha cell preserves its overall shape. Since we assume unlimited cell growth we describe steady tip growth by travelling wave profiles. For this purpose, we first need to introduce suitable coordinates to describe the cell surface.


\subsubsection{Parametrizing the hypha cell surface}

\begin{figure}[h]
\begin{center}
\includegraphics[width=8cm]{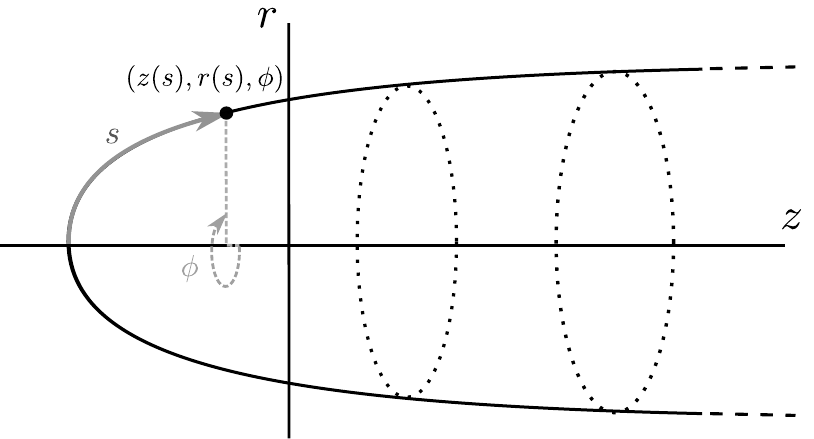}
\caption{The hypha cell shape: We assume that the cell shape is axially symmetric. Hence, we use cylindrical coordinates $(z(s),r(s),\phi)$ where $s$ is the arclength to the tip.  \label{fig:mshape}}
\end{center} \end{figure}

We assume  that the cell surface is axially symmetric. Hence, it is convenient to express the cell shape in cylindrical variables $(z,r,\phi)$. We parametrise the $z,r$-variables with respect to $s$ which is the arc length to the tip, $s=0$. This implies that the tip of the cell is the intersection of the cell surface with the $z$-axis. Hence, $\lim_{s \rightarrow 0}(r(s),z(s)) =(0,z_0)$. Observe that we have the equality
\begin{align*}
\left(  \frac{dr}{ds}\right)^2 + \left(  \frac{dz}{ds}\right)^2 =1. 
\end{align*}
Hence, $z(s)$ fixes the cell surface. For a visualization of the variables describing the cell shape see Figure \ref{fig:mshape}.\\

%

\subsubsection{Steady tip growth shape conditions}

We will characterize the cell shape during steady tip growth. We assume that $z,r \in C^{\infty}(\mathbb{R}_+, \mathbb{R})$. The following four conditions give a formal description of the steady tip growth shape:

\begin{itemize}
\item[T1a] \textbf{Tip limits:} 
\begin{align*}
\lim_{s \rightarrow 0} \frac{dr}{ds}(s)=1, \;  \;  \lim_{s \rightarrow 0} r(s) =0, \; \; \lim_{s \rightarrow 0}  z(s)= z_0 . 
\end{align*}
\item[T2a] \textbf{Analyticity in $r^2$:} There exists a $s_0>0$ and a $G_1 \in C^{\omega}\left( (-a^2 ,a^2) , \mathbb{R} \right)$ with $a=r(s_0)$ such that
\begin{align*}
z(s)= G_1(r(s)^2) \qquad \forall s \in (0,s_0).
\end{align*}
\item[T3a] \textbf{Global constraints:} For all $s \in \mathbb{R}_+$ the following constraints are satisfied
\begin{align*}
\frac{d^2r}{ds^2}(s) <0, \; \frac{dr}{ds}(s) >0, \; \frac{dz}{ds}(s) >0 .
\end{align*}
\item[T4a] \textbf{Base limits:}  
\begin{align*}
\lim_{s \rightarrow \infty}  r(s) =r_\infty>0,  \; \; \lim_{s \rightarrow \infty}  z(s)= \infty. 
\end{align*}
\end{itemize}
We refer to the conditions T1a, T2a, T3a, T4a as the steady tip growth shape conditions. We briefly explain these conditions.  By the definition of $s$ it follows that $\lim_{s \rightarrow 0}(z,r)(s)$ corresponds to the cell's tip. This explains condition T1a.  Condition T2a is based on the axial symmetry and the smoothness at the tip. We assume that the cell qualitatively resembles the cell shape in Figure \ref{fig:mshape} which explains condition T3a. We have that $\lim_{s \rightarrow \infty}(z,r)(s)$ corresponds to the cell's base, i.e. the base of the dome forming the apical shape. For tip growth we expect that the cell's width at the base converges to a positive constant. Hence, we require $\lim_{s \rightarrow \infty} r(s) = r_{\infty}>0$. In this idealized setting we assume that the hypha has infinite length. Hence, from the cell's orientation it follows that $\lim_{s \rightarrow \infty} z(s) =  \infty$. This explains condition T4a. 

\subsubsection{Travelling wave profiles}

\begin{figure}[h]
\begin{center}
\includegraphics[width=9cm]{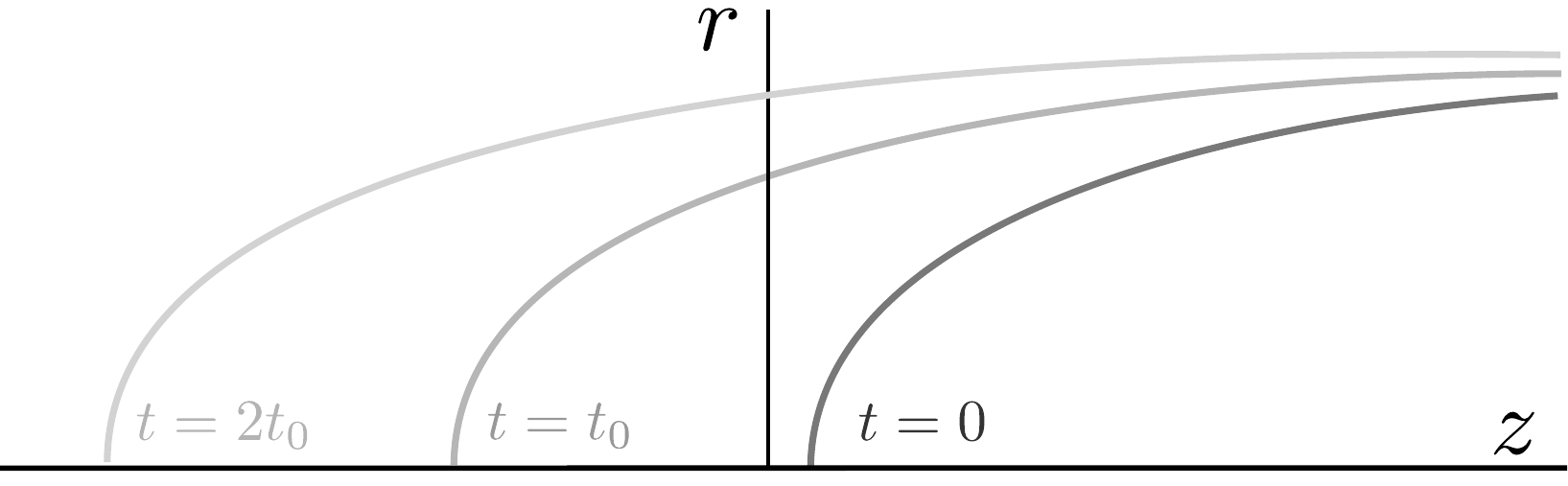}
\caption{Travelling profiles corresponding to steady tip growth: The tip moves from right to left in forward time. We displayed three profiles with timestep $t_0>0$. \label{travi}}
\end{center} \end{figure}

The hypha moves at an approximately constant speed and the hypha cell preserves its overall shape. The cell grows in the direction of the outward normal at the tip. We assume that the tip moves at an approximately constants speed and that the hypha cell preserves its overall shape. Let $z,r$ satisfy the steady tip growth shape conditions. Then, in the $(z,r)$-plane the moving profile at time $t$ is characterized by $(z(s)+ct,r(s))$ where $c$ is the velocity of the tip, see Figure \ref{travi}.

\subsection{The cell wall as a thin viscous sheet \label{sec:CM}}

Following Campas and Mahadevan \cite{CM09}, we model the cell wall as a thin viscous sheet subject to a pressure difference.    Let us first explain these modelling assumptions:
\begin{itemize}
\item[-] \textbf{The pressure difference:}  In the hypha biology this  is the difference between the turgor and the atmospheric pressure.  It generates a force in the direction of the outward normal since the turgor pressure is higher than the atmospheric pressure. 
\item[-] \textbf{Thin sheet:} It makes sense geometrically to model the cell wall as a thin sheet since the cell wall thickness is much smaller than the observed radii of curvature of the cell.  
\item[-] \textbf{The cell wall is viscous:} Little is known about the mechanical properties of the cell wall. However, the numerical work performed for the thin viscous sheet model of Campas and Mahadevan suggests the existence of solutions satisfying the steady tip growth shape conditions \cite{CM09}. 
\end{itemize}

To derive the governing equations for our tip growth model we will first assume that $z,r$ satisfy the steady tip growth shape conditions. In this section we will give a derivation of the governing equation when the cell wall is modelled as a thin viscous sheet subject to a pressure difference. We will also discuss conditions on the new variables in the case of steady tip growth. 

\subsubsection{Derivation of the thin viscous sheet equations}

We assume that the cell wall is a thin viscous sheet in mechanical equilibrium sustained by a pressure difference. Chapter 7 of Howell's doctoral thesis \cite{HW94} contains a derivation of the generalized thin viscous sheet equations. We use this to derive the thin viscous sheet equations for our axially symmetric cell surface. In addition, we include the pressures applied to the boundary of the sheet \textit{a priori}. In Howell's work the pressure drop is only included after derivation of the governing equations. Furthermore, in preparation of the Spitzenk\"{o}rper model, Section \ref{sec:BG}, we will assume that there is an influx of viscous fluid at the inner boundary of the sheet. \\

\begin{figure}[h]
\begin{center}
\includegraphics[width=6cm]{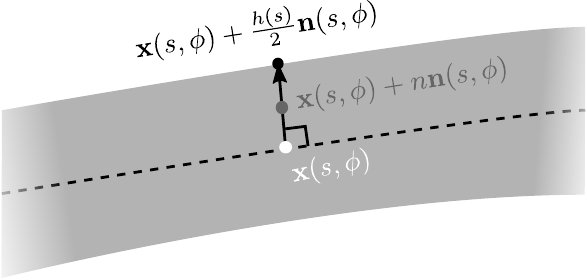}
\caption{Cell wall: The grey area represents part of the cell wall. The dashed line is the center surface. The cell wall is parameterised by $\textbf{x}(s,\phi) \pm n \mathbf{n}(s,\phi)$. At $s$ the cell wall has thickness $h(s)$.  \label{fig:xnh}}\end{center} \end{figure}

\textit{Parametrising the cell wall:}    Observe that in $\mathbb{R}^3$ any axially symmetric surface can be described by
\begin{align}
\textbf{x}(s, \phi)= \begin{bmatrix}
r(s) \cos (\phi) \\
r(s) \sin (\phi) \\
z(s)
\end{bmatrix} . \label{xvector}
\end{align} 
We let $\mathbf{x}$ given by (\ref{xvector}) parametrise the center surface of the sheet. Then by axial symmetry the cell wall thickness, $h$, is a function of only $s$.  The boundary of the cell wall is parametrized by $\textbf{x}(s,\phi) \pm \frac{h(s)}{2} \mathbf{n}(s,\phi)$, where $\mathbf{n}$ is the normal given by
\begin{equation}
\mathbf{n}(s,\phi) =   \frac{ \frac{d \textbf{x}}{d \phi} \times \frac{d \textbf{x}}{d s}}{|  \frac{d \textbf{x}}{d \phi} \times \frac{d \textbf{x}}{d s}|} = \begin{bmatrix}
\frac{dz}{ds}(s) \cos(\phi) \\
\frac{dz}{ds}(s) \sin(\phi) \\
-\frac{dr}{ds}(s) \\
\end{bmatrix}.
\end{equation}  
The position of a fluid particle in the sheet is represented by $\textbf{x}(s,\phi) + n \mathbf{n}(s,\phi) $ where $n \in (-h(s)/2,h(s)/2)  $. Observe that since our description of the position of the fluid particle is independent of time we are setting up governing equations in a moving coordinate frame.  We will refer to the normal direction as the $n$-direction. For an overview of the description of the cell wall see Figure \ref{fig:xnh}.  
The normal on the boundary $n=\pm h/2$ is given by 
\begin{align}
\mathbf{n}_{\pm}:=  {\rm grad} (-n \pm h /2 ) \big|_{n= \pm h/2} \label{nboundary}
\end{align}

\begin{figure}[h]
\begin{center}
\includegraphics[width=6cm]{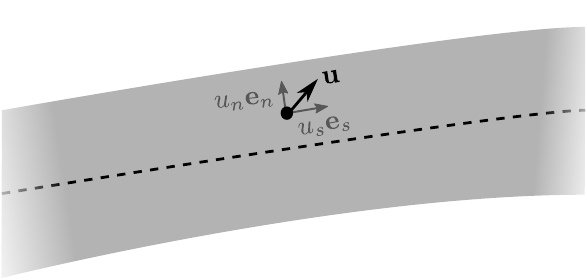}
\caption{The velocity vector $\mathbf{u}$: The dashed line is the center surface. Since the cell does not rotate we have that $\mathbf{u} = u_s \mathbf{e}_s + u_n \mathbf{e}_n$.  \label{fig:justuv}}
\end{center} \end{figure}

\textit{Incompressibility condition:} It will be convenient to use the orthonormal basis $\{ \mathbf{e}_s, \mathbf{e}_\phi, \mathbf{n} \}$ where $\mathbf{e}_s, \mathbf{e}_\phi$ correspond to the unit vectors in the $s$-, $\phi$-direction, respectively. The velocity vector of a fluid particle, $\mathbf{u}$, is a function of $(s,n)$. We can write   $\mathbf{u} = u_s \mathbf{e}_s +  u_\phi \mathbf{e}_\phi + u_n \mathbf{e}_n$. Since the cell does not rotate we assume that $u_\phi=0$. The $\mathbf{u}$ vector is visualised in Figure \ref{fig:justuv}.

Observe that $u_s,u_n$ are functions of $(s,n)$. The incompressibility condition  is  given by
\begin{align}
\frac{d( \ell_\phi u_s)}{d s} + \frac{d (\ell_s \ell_\phi u_n )}{d n} =0 ,  \label{incompres}
\end{align}
where $\ell_s, \ell_\phi$ correspond to scaling factors given by
\begin{align*}
\ell_s = 1- \kappa_s n , \qquad \ell_\phi = r (1- \kappa_\phi n ),
\end{align*}
and $\kappa_s, \kappa_\phi$ are the principal curvatures given by
\begin{align}
\kappa_s = \frac{- d^2 r/ d s^2}{{d z}/{d s}},   \qquad \kappa_\phi =  \frac{ d z/ d s}{r}. \label{princpsphicurv}
\end{align}

\textit{Stress-strain constitutive relation and Stokes equation:} Before we present the Stokes equation we present the linear stress-strain constitutive relation given by
\begin{gather}
\begin{aligned}
\sigma_{ss} &= - p + \frac{2 \mu_0}{\ell_s} \left( \frac{d u_s}{d s}  + u_n \frac{d \ell_s}{d n} \right), & \sigma_{\phi\phi} &= - p + \frac{2 \mu_0}{\ell_\phi} \left( \frac{u_s}{\ell_s}\frac{d \ell_\phi}{d s}  + u_n \frac{d \ell_\phi}{d n} \right), & \\
\sigma_{s\phi}& =0,  &
\sigma_{nn} &= -p + 2 \mu_0 \frac{d u_n}{d n},& \\
\sigma_{s n} &= \frac{\mu_0}{\ell_s} \left(  \ell_s \frac{d u_s}{d n}  -  \frac{d \ell_s}{d n} u_s + \frac{d u_n}{ds}\right), & 
\sigma_{\phi n} &=  0 ,&
\end{aligned} \label{stresscompp}
\end{gather}
where $\mu_0$ is the viscosity and $p$ is the hydrodynamics pressure inside the sheet. We assume that the viscosity $\mu_0$ is a function of $s$.  The Stokes equation in the $\phi$-direction yields the zero identity. The Stokes equation in the $s$-  and $n$-direction are  given by
\begin{align}
\frac{d (\ell_{\phi} \sigma_{ss})}{d s} + \frac{d (\ell_{s} \ell_{\phi} \sigma_{sn})}{d n} +  \ell_{\phi} \frac{d \ell_{s} }{d n}  \sigma_{sn} -  \frac{d \ell_{\phi} }{d s}  \sigma_{\phi \phi} &=0, \label{sstokes} \\
\frac{d (\ell_{\phi} \sigma_{s n}) }{d s} + \frac{d (\ell_{s} \ell_{\phi} \sigma_{n n}) }{d n} - \ell_{\phi } \frac{d \ell_{s}  }{d n}  \sigma_{ss}  - \ell_{s} \frac{d \ell_{\phi}  }{d n}  \sigma_{\phi \phi}   &= 0 . \label{nstokes}
\end{align}

\begin{figure}[h]
\begin{center}
\includegraphics[width=6cm]{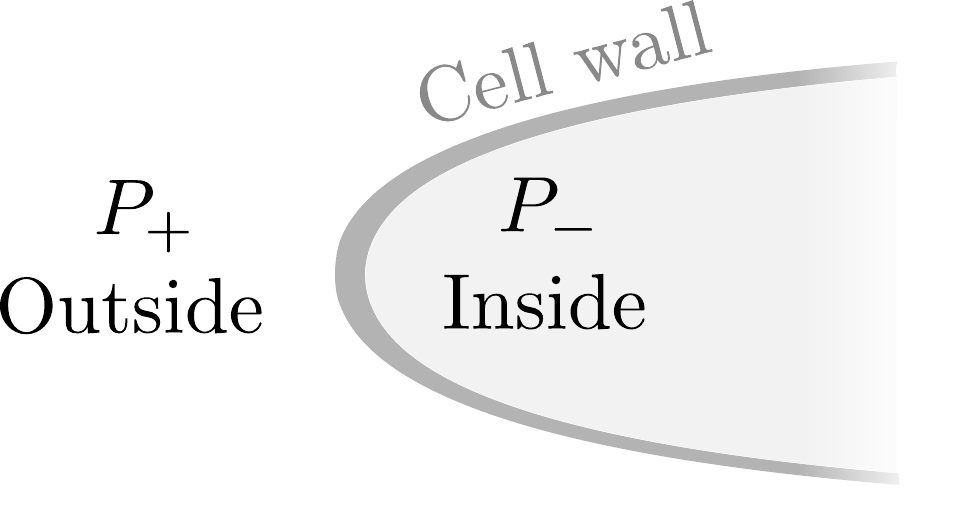}
\caption{Pressure difference: The light grey area corresponds to the inside of the cell and the dark grey area corresponds to the cell wall. The internal pressure is $P_-$ and the external pressure is $P_+$. \label{fig:Pdropdrop}}
\end{center} \end{figure}

\textit{Boundary conditions:} Denote by $P_{+}$ the pressure on the outside and by $P_{ -}$ the internal turgor pressure, see Figure \ref{fig:Pdropdrop}.

 Assume that $P_{ +} < P_{-}$.  The pressure difference produces a force in the normal direction on the sheet's boundaries. In the $s$-direction and the $\phi$-direction we assume that the external stresses are zero. Hence, we take as boundary condition $  \sigma \cdot \mathbf{n}_{\pm} =   P_{\pm}\mathbf{n}_{\pm}$ on  $n = \pm h/2$ where $\mathbf{n}_{\pm}$ is given by (\ref{nboundary}). Evaluating this boundary condition yields the following equalities on $n=\pm h/2$:
\begin{align}
\sigma_{s n} &=  \pm \frac{1}{2 \ell_s} \frac{d h}{d s} \left( \sigma_{ss} - P_{\pm} \right),   \label{stress0sn} \\
\sigma_{\phi n} &= \pm \frac{1}{2 \ell_s} \frac{d h}{d s} \sigma_{s \phi} \label{stressphn0}  \\
 \sigma_{n n} &= \pm \frac{1}{2 \ell_s} \frac{d h}{d s} \sigma_{sn} + P_{\pm}. \label{stress0nn}
\end{align} 
Observe that from (\ref{stresscompp}) it follows that (\ref{stressphn0}) does not contain any information. \\

There is an influx of new cell wall material through the inner surface boundaries. We denote the rate of cell wall addition per unit area by $\varphi$. Hence, we refer to $\varphi$ as the flux.
The kinematic condition   on $n=-h/2$ is then given by  $  \mathbf{u} \cdot \mathbf{n}_{-}= \varphi$. Evaluating this boundary condition yields the following equality on $n=-h/2$:
\begin{align}
 u_{n} =  - \frac{u_s}{2 \ell_{s}}\frac{d h}{d s } +  \varphi. \label{kinematic1}
\end{align}
Since there is no influx of new cell wall material at $n=h/2$ the kinematic condition on $n = h/2$ is given by
\begin{align}
 u_{n} =  \frac{u_s}{2 \ell_{s}}\frac{d h}{d s }. \label{kinematic2}
\end{align}

\textit{Non-dimensionalisation:} We will perform a non-dimensionalisation of the equations. Here we choose the length scale $L$ such that the characteristic mean curvature scale is of order $1/L$. We let the characteristic thickness be $\varepsilon L$ with $\varepsilon \ll 1$ a positive parameter. Denote by $U$ the characteristic velocity scale and by $M$ the characteristic viscosity scale. The scalings become    
\begin{align*}
r &= L r',&  s& = L s',& n &= \varepsilon L n',&  \\
h &= \varepsilon L h',& u_s& = U u_s',& u_n& = U u_n',& \\ 
\kappa_s &= \kappa_s'/L ,& \kappa_\phi &= \kappa_\phi'/L , &  \mu_0 &= M \mu_0',&\\
  p &= MU L^{-1} p'.&  
\end{align*}
The pressure, $P_{\pm}$, applies stress on a sheet with thickness of order $\varepsilon$. The flux $\varphi$ supplies new cell wall material to a sheet with thickness of order $\varepsilon$. Consequently, we assume that
\begin{align}
 P_{\pm} &=  \varepsilon MU L^{-1} P_{\pm}'  &    \varphi &= \varepsilon U L \varphi' .     
\end{align}

In the primed variables we can non-dimensionalise the governing equations. We continue with the primed variables. For notational we will drop the primes. 

\textit{Formal $\varepsilon$-expansions:} Assume that we can write the variables $u_s,u_n,p$ as a formal expansion in $\varepsilon$. We will denote the $i$th coefficient with a $(i)$ superscript where $i \in \mathbb{N}_{\geq 0}$, e.g. 
\begin{gather} 
\begin{aligned}
u_s&=u_s^{(0)} +  \varepsilon u_s^{(1)} + O(\varepsilon^2).
\end{aligned} \label{epsilonthings}
\end{gather}

Evaluating the $\varepsilon$-expansion in (\ref{stresscompp}) we obtain
\begin{gather}
\begin{aligned}
\sigma_{ss} &=    O(1), \\
\sigma_{\phi \phi}  &= O(1),         \\
\sigma_{n n} &=\frac{ 2 \mu_0}{\varepsilon} \frac{d u_n^{(0)}}{d n} + O(1), \\
\sigma_{s n} &= \frac{ \mu_0}{\varepsilon} \frac{d u_s^{(0)}}{dn } + O(1)
\end{aligned} 
\end{gather}
Then collecting the leading-order terms in the Stokes equation and the zero stress boundary condition we obtain that
\begin{align*}
\frac{d u_n^{(0)}}{d n}=0, \qquad \frac{d u_s^{(0)}}{dn } =0.
\end{align*}
Hence, $u_n^{(0)}$ and $u_s^{(0)}$ are independent of $n$ and  the leading order terms of the stress components $\sigma_{nn}$ and $\sigma_{sn}$ are $O(1)$.  We denote the $i$th coefficient of the stress components with a $(i)$ superscript where $i \in \mathbb{N}_{\geq 0}$. \\

 Collecting the leading order terms in the kinematic condition and using that $u_n^{(0)}$ is independent of $n$ gives
\begin{align}
u_n^{(0)} =0 . \label{un0}
\end{align}
Using (\ref{un0}) the leading order terms of the stress components become
\begin{gather}
\begin{aligned}
\sigma_{ss}^{(0)} &= -p^{(0)}  + 2 \mu_0 \frac{d u^{(0)}_s}{d s}    ,         \\
\sigma_{\phi \phi}^{(0)} & = -p^{(0)}  + \frac{2 \mu_0  u_s^{(0)} }{r}    \frac{d r}{d s}  , \\
\sigma_{n n}^{(0)} &=  -p^{(0)}  + 2 \mu_0 \frac{d u^{(1)}_n}{d n} , \\
\sigma_{s n}^{(0)} &= \mu_0\left(  \frac{d u^{(1)}_s}{d n} + \kappa_s  u^{(0)}_s  \right)  .
\end{aligned} \label{stress0}
\end{gather}
Collecting the lowest order terms in $\varepsilon$ in the incompressibility conditions we get the equality
\begin{align}
\frac{d u_{n}^{(1)}}{d n} = -\frac{1}{r} \frac{d (r u_s^{(0)})}{d s} . \label{un1}
\end{align}
Hence, $u_{n}^{(1)}$ is linear in $n$. From the kinematic condition we obtain that
\begin{align}
 u_n^{(1)} &=\frac{u_s^{(0)}}{ 2} \frac{d h}{d s}  &{\rm on }& \; \; \; n=h/2 , \label{un2}\\
 u_n^{(1)} &= -\frac{u_s^{(0)}}{2 } \frac{d h}{d s} + \varphi^{(0)}  &{\rm on }& \; \;  \; n=-h/2.  \label{un3}
\end{align}
Combining (\ref{un1})-(\ref{un3}) we obtain the equality
\begin{align*}
- \frac{h}{r} \frac{dr u_s^{(0)}}{ds} =  u_s^{(0)} \frac{dh}{ds} - \varphi^{(0)}. 
\end{align*}
This equality can be written as 
\begin{align}
\frac{d (r h  u_s^{(0)} )}{d s} = r \varphi^{(0)}. \label{prevol} 
\end{align}
Collecting the leading order terms in the Stokes equation  (\ref{stress0nn}) and  in the zero stress condition (\ref{stress0nn}) we get that $\sigma_{nn}^{(0)} =0$. Consequently, the leading order pressure term becomes \begin{align}
p^{(0)} = 2 \mu_0 \frac{ d u_{n}^{(1)}}{d n} . \label{p0}
\end{align}
Inserting (\ref{un1}) into  (\ref{p0})  we obtain
\begin{align}
p^{(0)} = - \frac{2 \mu_0^{(0)}}{r}    \frac{d (r u_s^{(0)})}{d s} . \label{p0e}
\end{align}
Inserting (\ref{p0e}) into the equation for $\sigma_{ss}$  and $\sigma_{\phi \phi}$ (\ref{stress0}) we obtain
\begin{gather}
\begin{aligned}
\sigma_{ss}^{(0)} &=4 \mu_0\left( \frac{d u^{(0)}_s}{d s} + \frac{ u^{(0)}_s }{2 r} \frac{d r}{d s}  \right) ,\\
\sigma_{\phi \phi}^{(0)} &= 4 \mu_0\left( \frac{1}{2} \frac{d u^{(0)}_s}{d s} + \frac{u^{(0)}_s  }{ r}  \frac{d r}{d s} \right) .
\end{aligned} \label{sigmassphiphi}
\end{gather}
Hence, the leading order terms of $\sigma_{ss}$  and $\sigma_{\phi \phi}$ are independent of $n$. Collecting the leading order terms in the Stokes equation (\ref{sstokes}) and zero stress condition  (\ref{stress0sn}) we obtain that 
\begin{align}
\sigma_{sn}^{(0)} = 0. \label{stsn0}
\end{align}
Consequently,  collecting the leading order terms in the Stokes equation (\ref{nstokes}) and using (\ref{stsn0}) gives
\begin{align}
\frac{d \sigma_{nn}^{(1)}}{d n} + \kappa_s \sigma_{ss}^{(0)} + \kappa_\phi \sigma_{\phi \phi}^{(0)} =0. \label{nstressblanace}
\end{align}
Hence, $ \sigma_{nn}^{(1)}$ is linear in $n$. Collecting the leading order terms of the zero stress condition (\ref{stressphn0}) and using (\ref{stsn0}) we get
\begin{align}
 \sigma_{nn}^{(1)}  =  P_{\pm}  \qquad {\rm on} \;\; n= \pm h/2. \label{snnbound}
\end{align} 
\textit{Cell wall tension equations:} We define the pressure drop $P:=P_- - P_+$. Tension in a surface is the force across an infinitesimal path per unit length. Hence, tension is an analogue of stress for surfaces.  The leading order terms of the tensions on the sheets' boundary in the $s$- and $\phi$-direction are given by $\overline{\sigma}_{ss}:=  h\sigma_{ss}^{(0)}$ and  $\overline{\sigma}_{\phi \phi}:=  h \sigma_{\phi \phi}^{(0)}$, respectively. Combining (\ref{nstressblanace}) and (\ref{snnbound}) we obtain
\begin{align}
 \kappa_{s} \overline{\sigma}_{ss} +  \kappa_\phi \overline{\sigma}_{\phi \phi} = P .  \label{firstsspr}
\end{align}
Collecting the leading order terms of the Stokes equation (\ref{sstokes}) we get
\begin{align}
\frac{d (r \sigma_{ss}^{(0)})}{d s} + \frac{d (r \sigma_{sn}^{(1)})}{d n} - \frac{d r}{d s} \sigma_{\phi \phi}^{(0)} =0 . \label{sstressblanace}
\end{align}
Hence, $\sigma_{sn}^{(1)}$ is linear in $n$. Collecting the leading order terms of the zero stress condition (\ref{stress0sn}) we get 
\begin{align}
\sigma_{sn}^{(1)} &= \pm \frac{1}{2} \frac{dh}{ds}  \sigma_{ss}^{(0)} \qquad {\rm on} \;\; n= \pm h/2. \label{sigmasnnnr}
\end{align}
Combining (\ref{sstressblanace}) and (\ref{sigmasnnnr}) we get
\begin{align}
r \frac{dh}{ds} \sigma_{ss}^{(0)}  =  h  \left( \frac{dr}{ds} \sigma_{\phi \phi}^{(0)} - \frac{d(r \sigma_{ss}^{(0)})}{ds} \right). \label{sstressbalancennnr}
\end{align}
This equality can be written as 
\begin{align}
\frac{d (r  \overline{\sigma}_{ss})}{d s} - \frac{d r}{d s} \overline{\sigma}_{\phi \phi} =0 .\label{firststry}
\end{align}
Using the expression for the curvature terms and (\ref{firstsspr}) we can rewrite (\ref{firststry}) in a nicer form \cite{PIP68} (see also p. 212 in \cite{GO17}):
\begin{align*}
\frac{d (r  \overline{\sigma}_{ss})}{d s} &= \frac{d r}{d s} \overline{\sigma}_{\phi \phi} , \\
\frac{d z}{d s} \frac{d (r  \overline{\sigma}_{ss})}{d s}  &= r \frac{ d r }{d s} \left( P  - \frac{d^2 z/d s^2}{d r/ d s} \overline{\sigma}_{ss}\right),\\
\frac{d z}{d s} \frac{d (r  \overline{\sigma}_{ss})}{d s} + r \frac{d^2 z}{d s^2} \overline{\sigma}_{ss} &= r \frac{ d r }{d s}  P   ,\\
 \frac{d (r^2 \kappa_{\phi}  \overline{\sigma}_{ss})}{d s} &= r \frac{ d r }{d s}  P , \\
\int^{s}_0  \left(\frac{d (r^2 \kappa_{\phi}  \overline{\sigma}_{ss})}{d \tau} \right) (\tau) d \tau &= P \int^{s}_0   r(\tau) \frac{d r }{d \tau}(\tau)   d \tau ,\\  
\left[ \left( r^2 \kappa_{\phi}  \overline{\sigma}_{ss} \right) (\tau) \right]_{0}^s  & = \left[ \frac{P}{2} r(\tau)^2 \right]_{0}^{s} . 
\end{align*}
Assuming that $r,z$ satisfy the tip limits T1a we get
\begin{align*}
\kappa_{\phi} \overline{\sigma}_{ss}= \frac{P}{2}. 
\end{align*}

\textit{Equations for the cell wall:} We continue with the leading order terms. Hence, we drop all the superscripts. Let $u := u_s$. Then to summarize this section we have the following equations for the cell wall:
\begin{align}
 \kappa_{s} \overline{\sigma}_{ss} +  \kappa_\phi \overline{\sigma}_{\phi \phi} &= P ,  \label{normalstressf} \\ 
\kappa_{\phi} \overline{\sigma}_{ss}&= \frac{P}{2},  \label{sstresscurf} \\ 
\frac{d}{ds} \left(r h  u \right) &= r \varphi,  \label{prevol2}
\end{align}
where the tensions are given by
\begin{gather}
\begin{aligned}
\overline{\sigma}_{ss}  &=  4 \mu_0 h \left( \frac{d u}{d s} + \frac{ u }{2 r} \frac{d r}{d s}  \right), \\
\overline{\sigma}_{\phi \phi} & = 4 \mu_0 h \left( \frac{1}{2} \frac{du}{ds} + \frac{u  }{ r}  \frac{d r}{d s} \right) .
\end{aligned} \label{sumtt} 
\end{gather}
Observe that all the equations are independent of time since we gave a description of the cell wall in comoving frame which moves with the travelling wave profiles' velocity. The velocity of the travelling wave profile can be obtained by computing $\lim_{s \rightarrow \infty} u(s)$.\\

In the supplemental data of \cite{CM09} Campas and Mahadevan give a heuristic derivation of (\ref{normalstressf})-(\ref{sstresscurf}). Equation (\ref{normalstressf}) and (\ref{sstresscurf}) are derived by assuming a force balance in the $n$- and $z$-direction, respectively. Hence, we refer to equation (\ref{normalstressf}) and (\ref{sstresscurf}) as the $n$- and $z$-balance equation, respectively. We refer to equation (\ref{prevol2}) as the mass balance equation since it is derived from assuming a mass balance.

\subsubsection{Steady tip growth dynamics conditions}

 Similarly to what we did for the shape variables, $r,z$, we  formulate steady tip growth conditions for $u,h$:

\begin{itemize}
\item[T1b] \textbf{Tip limits:} 
\begin{align*}
\lim_{s \rightarrow 0} u(s)=u_0>0, \qquad \lim_{s \rightarrow 0} h(s)=h_0>0. 
\end{align*}
\item[T2b] \textbf{Analyticity in $r^2$:} There exists a $s_0>0$ and a $G_2 \in C^{\omega}\left( (-a^2 ,a^2) , \mathbb{R} ^2\right)$ with $a=r(s_0)$ such that
\begin{align*}
(u(s),h(s))= G_2(r(s)^2) \qquad \forall s \in (0,s_0).
\end{align*}
\item[T3b] \textbf{Global constraints:} For all $s \in \mathbb{R}_+$ the following constraints are satisfied:
\begin{align*}
u(s)>0, \qquad h(s)>0.
\end{align*}
\item[T4b] \textbf{Base limits:}  
\begin{align*}
\lim_{s \rightarrow \infty}  u(s) =u_\infty>0,  \; \; \lim_{s \rightarrow \infty}  h(s)=  h_\infty>0. 
\end{align*}
\end{itemize}

We refer to the conditions T1b, T2b, T3b, T4b as the steady tip growth dynamics conditions. We briefly explain these conditions. The cell wall particles cannot be stationary, as otherwise the cell wall would continuously thicken. We also expect that $\lim_{s \rightarrow \infty} u(s)= u_{\infty}>0$ since the velocity of the  steady tip growth profile must be a positive constant. The cell wall thickness must be positive for steady tip growth. This explains condition T1b, T3b, T4b. Condition T2b is based on the smoothness and axial symmetry of the cell shape.\\

\subsection{The Spitzenk\"{o}rper as a ballistic point source \label{sec:BG}}

\begin{figure}[h]
\begin{center}
\includegraphics[width=8cm]{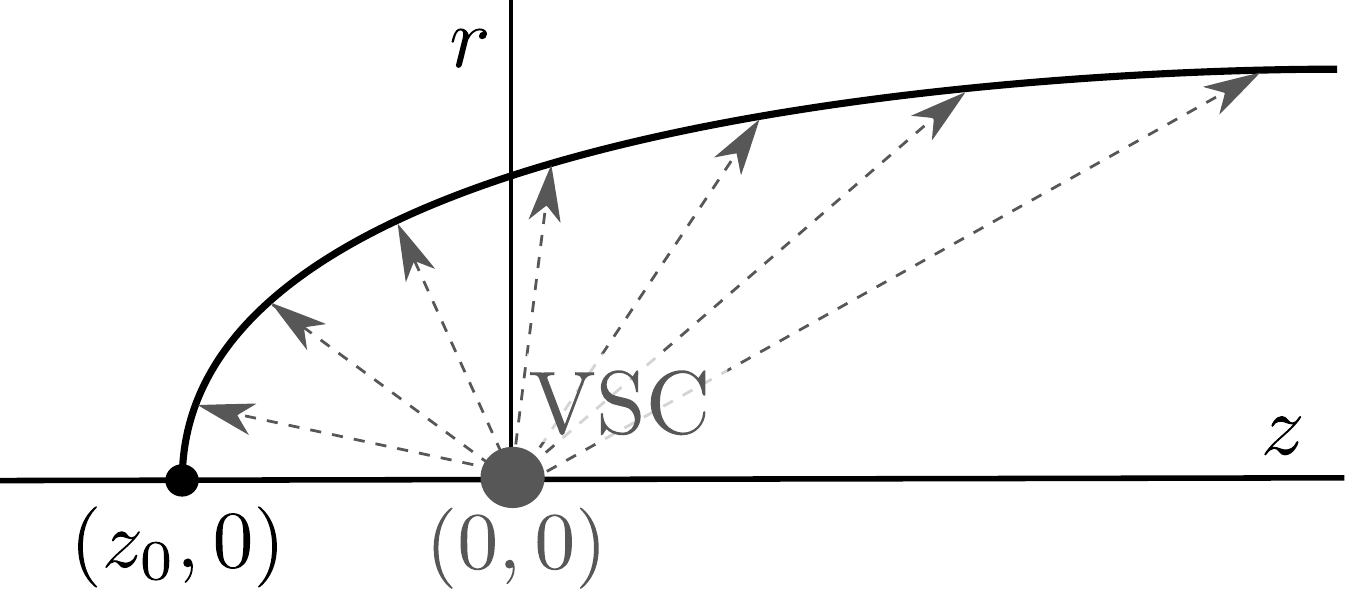}
\caption{Ballistic vesicle supply center: the VSC is a point source at $(0,0)$. It is located at a fixed distance from the tip, $(z_0,0)$ with $z_0<0$. The ballistic VSC sends vesicles in a straight line to the cell wall. \label{fig:bali}}\end{center} \end{figure}

We model the Spitzenk\"{o}rper using the work of Bartnicki-Garcia et al. \cite{BG89}. They assume that the Spitzenk\"{o}rper can be modelled as a  so-called Vesicle Supply Center (VSC). They take the VSC to be an isotropic point source which continuously and at a constant rate sends vesicles to the cell wall. Furthermore, they assume that it transports vesicles in straight lines to the cell wall, see Figure \ref{fig:bali}. Therefore, it is referred to as the ballistic VSC.  \\

\begin{figure}[h]
\begin{center}
\includegraphics[width=6cm]{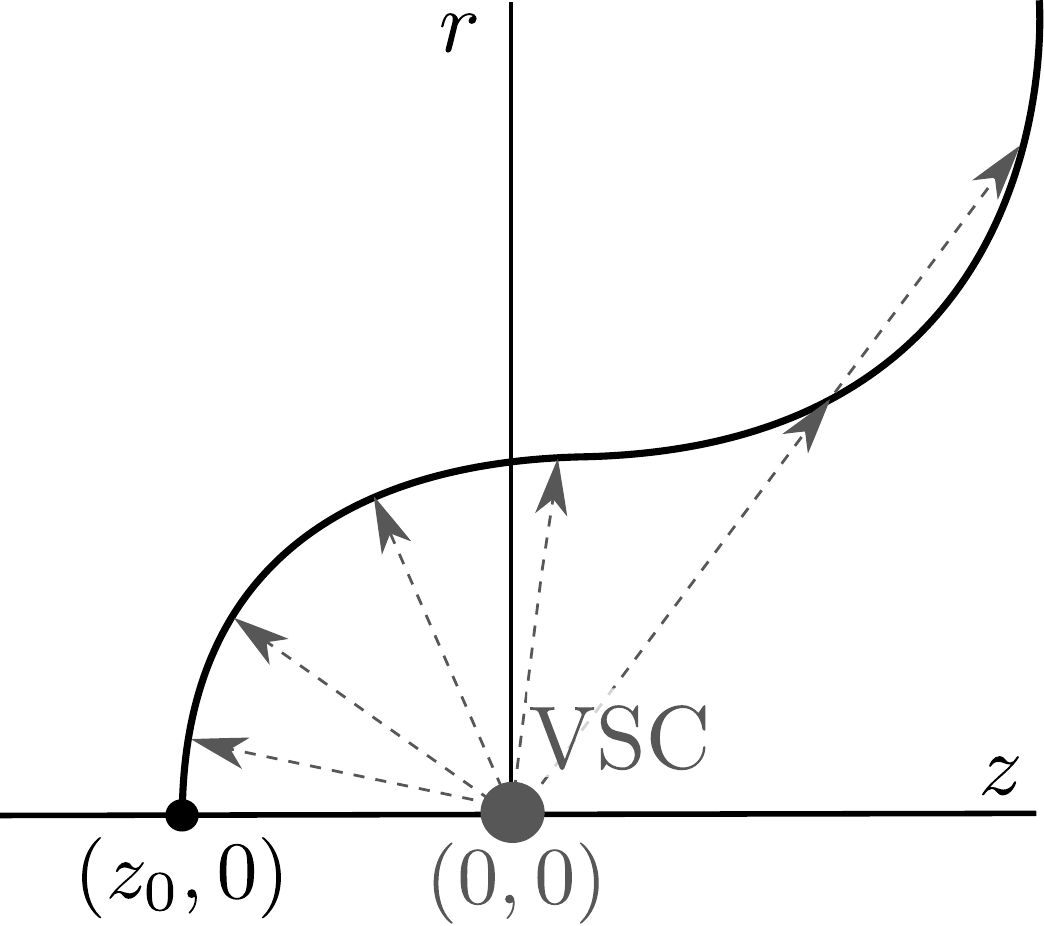}
\caption{Example of a  non-star-shaped cell. Observe that  the cell wall shape does not satisfy T3a. \label{fig:notcbali}}\end{center} \end{figure}

We place the VSC at the origin in the $(z,r)$-plane. We assume that $z,r$ satisfy the steady tip growth shape conditions with $\lim_{s \rightarrow 0} z(s) = z_0<0$, see Figure \ref{fig:bali}. For this model to make sense, the cell needs to be star-shaped with respect to the VSC, see Figure \ref{fig:notcbali}. Observe that this is guaranteed by Condition T3a.

Based on the modelling considerations we derive the vesicle flux at the internal boundary of the cell wall. The scalar flux at a point on the cell wall is given by     
\begin{align*}
\frac{C_0}{4 \pi (r^2 + z^2)^{3/2}} \textbf{x}(s,\phi) \cdot \mathbf{n}(s,\phi) =  \frac{C_0}{4\pi} \frac{r \cdot dz/ds-z \cdot dr/ds}{(z^2+r^2)^{3/2}} ,
\end{align*}
here $C_0$ is the rate of cell-wall building material emitted by the VSC.

Let $C = C_0/(4 \pi)$. By T3a we have  $dz/ds=\sqrt{1- (dr/ds)^2}$. We define  the scalar flux
\begin{align*}
\gamma(dr/ds,z,r)  := C \frac{r \cdot \sqrt{1-(dr/ds)^2}-z \cdot dr/ds}{(z^2+r^2)^{3/2}},
\end{align*} 
  and, accordingly, fix $\varphi= \gamma(dr/ds,z,r)$ in (\ref{prevol2}).

\subsection{The viscosity and age function \label{sec:va}}
Inserting the curvature terms (\ref{princpsphicurv}) and the tensions (\ref{sumtt}) in the $z$-balance equation (\ref{sstresscurf}) we get
\begin{align}
4 \mu_0 h \frac{dz/ds}{r} \left( \frac{d u}{d s} + \frac{ u }{2 r} \frac{d r}{d s}  \right)  &= \frac{P}{2}   . \label{thiscontral}
\end{align}
 Assume for the moment that $\mu_0$ is constant (contrary to  \cite{CM09}).  Suppose that $z,r$ satisfy the  steady tip growth shape conditions and that $u,h$ satisfy the dynamic conditions. Taking the limit $s \rightarrow \infty$ in (\ref{thiscontral}) yields 
\begin{align}
\frac{4 \mu_0 h_\infty}{r_\infty} \lim_{s \rightarrow \infty} u(s) = \frac{P}{2} >0, \label{contraaaaup}
\end{align}
and so contradicts Condition T4b. Concluding, if $\mu_0$ is constant there exist no solutions to the model which satisfy the steady tip growth shape and dynamics conditions. \\

Hence, we will consider non-constant viscosity $\mu_0$. More specifically, we will define a new $s$-dependent variable $\Psi$ corresponding to the average age of the cell wall material at $s$. Then we will set $\mu_0= \mu(\Psi)$ with $\mu \in C^{\infty}(\mathbb{R}_+, \mathbb{R}_+)$ a suitably chosen function. We refer to $\mu$ as the viscosity function. \\

Let us first set up the age equation. We write the position of a cell wall particle as $s(t)$ where $t$ is the time variable. Then we can write the tangential velocity of a cell wall particle as
\begin{equation}
u= \frac{ds}{dt} \label{veloctime}.
\end{equation}
Denote by $t(\varsigma,s)$ the time it takes a particle in the cell wall to travel from $\varsigma$ to $s$. Using (\ref{veloctime}) we have that 
\begin{equation}
t(\varsigma,s)= \int^{s}_\varsigma \frac{1}{u(\sigma)} d \sigma .
\end{equation}
We define the cumulative flux $G(s)$ as total flux of material into the part of the cell wall parameterized by arclength $\varsigma$ from the tip to $\varsigma=s$:
\begin{align*}
G(s) :=  \int^{s}_0 (r \gamma(dr/d\varsigma,z,r))(\varsigma)  d \varsigma    .
\end{align*}
At a point  with arclength $s$ there is  cell wall material which originally entered at $\varsigma \in(0,s)$ and has been part of the cell wall for $t(\varsigma,s)$. Weighing $t(\varsigma,s)$ by the mass which enters the cell wall at $\varsigma$ we get $t(\varsigma,s)G'(\varsigma)$.  To obtain the average age we  integrate $t(\varsigma,s)G'(\varsigma)$ over an arc $(0,s)$ and  divide it by the total  flux over  that arc:
\begin{equation}
\Psi(s) = \frac{ \int^{s}_0 G'(\varsigma) t(\varsigma, s)  d \varsigma  }{ G(s) } \label{aged} .
\end{equation}
We rewrite the expression for $\Psi$ into a more convenient form using the mass balance equations and $G$:
\begin{align}
\Psi(s) &= \frac{ \int^{s}_0 G'(\varsigma) t(\varsigma, s)    d \varsigma  }{ G(s) } ,\nonumber  \\
 &= \frac{ \int^{s}_0 \frac{G(\varsigma)}{u(\varsigma)}      d \varsigma  - \left[ G(\varsigma) \int^{s}_\varsigma \frac{1}{u(\sigma)} d \sigma  \right]_{\varsigma=0}^{\varsigma=s}}{ G(s) } , \nonumber \\
 &= \frac{ \int^{s}_0 (rh)(\varsigma)      d \varsigma }{ G(s) }.  \label{PsirhGamma}
\end{align}
Assume that $z,r$ satisfy the steady tip growth shape and that $u,h$ satisfy the dynamic conditions. Then $\Psi$ satisfies
\begin{align}
\Psi(s)>0  \;\;\;\;  \forall s \in (0, \infty), \;  \lim_{s \rightarrow 0} \Psi(s) = \frac{ h_0 z_0^2}{C }>0 ,  \; \lim_{s \rightarrow \infty} \Psi(s) = \infty . \label{Psicc}
\end{align} 
For the coupling between viscosity and age we introduce the function $\mu \in C^\infty(\mathbb{R}_+, \mathbb{R}_+)$ and set $\mu_0 = \mu(\Psi)$. We will assume that
\begin{gather}
\begin{aligned}
\mu'(\Psi) >0. \label{GC1}
\end{aligned}
\end{gather}
Observe that (\ref{GC1}) implements the soft spot hypothesis since the cell wall viscosity increases with respect to its average age. It then follows from (\ref{thiscontral}) by using  $\lim_{s \rightarrow \infty} \Psi(s) = \infty$  that
\begin{align}
\lim_{\Psi \rightarrow  \infty} \mu(\Psi) = \infty, \label{GC2}
\end{align}
is a necessary condition for the existence of a solution satisfying the steady tip growth shape and dynamics conditions. Hence, we will only consider viscosity functions $\mu$ which satisfy (\ref{GC2}).

\section{The governing first order ODE \label{secgov}}

We consider the dependent variables $r,z,h,u,\Psi$ and denote the $s$-derivative with a prime. The governing equations are then given by
\begin{align}
 \kappa_{s} \overline{\sigma}_{ss} +  \kappa_\phi  \overline{\sigma}_{\phi \phi} &= P \label{stressnnnn} \\
  \kappa_{\phi} \overline{\sigma}_{ss}&= \frac{P}{2}, \label{stresszzzz} \\
 \left(r h  u \right)' &= r \gamma(r',z,r) \label{massss} \\
   z'&=\sqrt{1 -r'^2}, \label{zzzz}
\end{align}
where
\begin{align}
\gamma(dr/ds,z,r) &= C \frac{r \cdot \sqrt{1-(dr/ds)^2}-z \cdot dr/ds}{(z^2+r^2)^{3/2}},
\end{align}
\begin{gather}
\begin{aligned}
\overline{\sigma}_{ss}  &=  {4 \mu(\Psi) h}\left(u' +  \frac{ u  r' }{2 r}   \right), &
\overline{\sigma}_{\phi \phi} & =  {4 \mu(\Psi) h} \left( \frac{u'}{2}  +  \frac{u  r' }{ r}   \right),& \\
\kappa_s &= - \frac{r''}{z'} ,&   \qquad \kappa_\phi &=  \frac{z'}{r},& 
\end{aligned} \label{thingsingovvv} 
\end{gather}
with
\begin{align}
\Psi(s) &= \frac{ \int^{s}_0 (rh)(\varsigma)      d \varsigma }{ G(s) }, & G(s) &:=  \int^{s}_0 (r \gamma(dr/d\varsigma,z,r))(\varsigma)  d \varsigma    .
  \label{againage}
\end{align}
The main problem is to find a solution of (\ref{stressnnnn})-(\ref{zzzz}) which satisfies the steady tip growth shape and dynamics conditions. In this section we reformulate this problem as finding a specific solution of a five dimensional first order autonomous ODE. First we perform a scaling to  eliminate an unessential parameter.

\subsection{Scaling}
 We define the following scaled functions and variables:
\begin{align*}
\tilde{\gamma}&= \gamma /C ,& \tilde{G} &= G/C,  \\
\tilde{h} &= h /C, &   \tilde{P} &= P /C.
\end{align*}
Accordingly, we define the scaled tensions:
\begin{gather*}
\begin{aligned}
\tilde{{\sigma}}_{ss}  &=   {4  \mu(\Psi) \tilde{h}} \left(u' +  \frac{ u  r' }{2 r}   \right), &
\tilde{{\sigma}}_{\phi \phi} & =  {4  \mu(\Psi)  \tilde{h}} \left( \frac{u'}{2}  +  \frac{u  r' }{ r}   \right).& 
\end{aligned} 
\end{gather*}
Then (\ref{stressnnnn})-(\ref{zzzz}) take the form
\begin{align*}
 \kappa_{s} \tilde{\sigma}_{ss} +  \kappa_\phi \tilde{\sigma}_{\phi \phi} &= \tilde{P} , &  \kappa_{\phi} \tilde{\sigma}_{ss}&= \frac{\tilde{P}}{2}, \\
 \left(r \tilde{h}  u \right)' &= r \tilde{\gamma}(r',z,r),  &  z'&=\sqrt{1 -r'^2},
\end{align*} 
and we can write the $\Psi$-equation (\ref{againage}) as
\begin{align*}
\Psi(s) &= \frac{ \int^{s}_0 (r\tilde{h})(\varsigma)      d \varsigma }{ \tilde{G}(s) }. \; \;  
\end{align*}
Observe that dropping the tildes yields the original equations.  This is because the involved quantities scale linearly with respect to the vesicle supply rate.  Consequently, without loss of generality we set $C=1$. 

\subsection{The cumulative flux as function of the shape \label{cumff}}

We will show that if $z,r$ satisfy the steady tip growth shape conditions then there exists a function $\Gamma \in C^{\infty}(\mathbb{R} \times \mathbb{R}_+, \mathbb{R}_+)$ such that $\Gamma(z(s),r(s))=G(s)$. \\

%

Using the substitution $v=z/r$ we find that 
\begin{align}
\int \left(r \gamma(r',z,r) \right)(\sigma) d \sigma =& \int   \frac{1}{(1+v^2)^{3/2}} d v ,  \label{cumagsub}
\end{align}
Since $z,r$ satisfy the steady tip growth shape conditions we have that 
\begin{align}
\lim_{s \rightarrow 0} \frac{z}{r}(s) = - \infty. \label{cumaglim}
\end{align}
Therefore, with  (\ref{cumagsub}) and (\ref{cumaglim})  the cumulative flux can be written as 
\begin{align*}
G(s) &= \int^s_0 r(\sigma) \gamma(r'(\sigma),z(\sigma),r(\sigma))  d \sigma 
 = \int^{z(s)/r(s)}_{-\infty}   \frac{1}{(1+v^2)^{3/2}} d v \\ 
&=\frac{z(s)/r(s)}{\sqrt{(z(s)/r(s))^2 + 1}} +1
=\frac{z(s)}{\sqrt{z(s)^2 + r(s)^2}} +1 .
\end{align*}
Consequently, we set
\begin{align}
\Gamma(z,r) := \frac{z}{\sqrt{z^2 + r^2}} +1. 
\end{align}
{red So, we will} substitute $G(s)$ by $\Gamma(z(s),z(s))$. Assuming that $u,h$ satisfy the steady tip growth dynamics condition we obtain from the mass balance equation (\ref{prevol2}): 
\begin{align}
u r h = \Gamma(z,r). \label{newmassb}
\end{align}

\subsection{ The differential equation for $\Psi$ \label{Psiid}}

The age variable $\Psi$ is defined by an integral equation, see (\ref{againage}). We will derive a differential equation for $\Psi$.\\

Observe that $\Psi$ satisfies 
\begin{equation}
\Psi' = \frac{rh}{\Gamma(z,r)} - \frac{r \gamma(r',z,r) }{\Gamma(z,r) } \Psi . \label{ODEage}
\end{equation}

 We replace the integral equation (\ref{againage}) by the differential equation (\ref{ODEage}). In contrast to (\ref{againage}), this differential equation does not imply the limit behavior of $\Psi$ near $s=0$. So to ensure that (\ref{Psicc}) is satisfied, we now have to require
 
\begin{align}
\lim_{s \rightarrow 0} \Psi(s) = h_0 z_0^2.  \label{psicondd}
\end{align}

\subsection{Eliminating the $u$ variable \label{decu}}

Now we will eliminate the $u$ variable from the governing equations and write the resulting equations in a convenient form.\\

Substituting the tensions  in (\ref{thingsingovvv}) in the balance equations (\ref{stressnnnn}), (\ref{stresszzzz}) we write the resulting equations as
\begin{align*}
 \begin{bmatrix}
\kappa_\phi & \kappa_\phi \frac{r'}{2r} \\
\kappa_s+\frac{\kappa_\phi}{2} & \frac{r'}{r} \left( \frac{\kappa_s}{2} + \kappa_\phi \right)
\end{bmatrix}
\begin{bmatrix}
u' \\
u
\end{bmatrix}
=\begin{bmatrix}
\frac{1}{2 \tilde{\mu}(\Psi) h} \\
\frac{1}{\tilde{\mu}(\Psi) h}
\end{bmatrix},
\end{align*}
where $\tilde{\mu}(\Psi) =  4 \mu(\Psi)/ P   $. Solving  for $u'$ we get
\begin{align}
u' = - \frac{r^2r''}{3  \tilde{\mu}(\Psi)z'^3 h  } . \label{duexp}
\end{align}

  Inserting  $u'$ from (\ref{duexp}) and $u$ from  (\ref{newmassb}) in the $z$-balance equation (\ref{stresszzzz})   we get
\begin{equation}
r'' = \frac{3}{2} \frac{z'^2}{r} \left(-1+  \frac{\Gamma(z,r) r' z' \tilde{\mu}(\Psi)}{r^3 }  \right) . \label{Drr}
\end{equation}
Hence, we obtained a second order equation for $r$. \\

Reinserting $r''$ from (\ref{Drr}) in  (\ref{duexp})   we get
\begin{align}
 u' = \frac{1}{2 h} \left( \frac{r }{z'\tilde{\mu}(\Psi)}- \frac{\Gamma(z,r) r' }{r^2} \right) .\label{anotherdu}
\end{align}

 From the mass balance equation (\ref{massss}) we get, using (\ref{newmassb}) and
(\ref{anotherdu}),
\begin{align}
h'&=\frac{1}{ru}(r\gamma(r',z,r)-r'hu-u'r h)\nonumber\\
&=\left( \frac{r \gamma(r',z,r)}{\Gamma(z,r)}- \frac{r'}{2r}- \frac{r^2 }{2 \Gamma(z,r) z' \tilde{\mu}(\Psi)}\right) h .
 \label{htequation}
\end{align}

 Observe that Equations (\ref{htequation}), (\ref{Drr}), as well as  (\ref{ODEage}) are independent of $u$.  Thus, we have eliminated the variable $u$ from the system. \\

For notational convenience we will drop the tilde on $\mu$.

\subsection{The five dimensional first order ODE}

We now summarize our results by presenting the governing equations as a first order ODE system. Since we want this subsection to be self-contained in defining this ODE we will also repeat all the necessary functions. The five dimensional first order autonomous ODE is given by
\begin{gather}
\left.
\begin{aligned}
\rho' &= \frac{3}{2} \frac{1-\rho^2}{r} \left(-1+  \frac{\Gamma(z,r) \mu(\Psi) \rho \sqrt{1-\rho^2}}{r^3 }  \right) , \\
 h'& = \left( \frac{r \gamma(\rho,z,r)}{\Gamma(z,r)}- \frac{\rho}{2r}- \frac{r^2 }{2 \Gamma(z,r) \mu(\Psi) \sqrt{1-\rho^2}}\right) h , \\
 \Psi' &= \frac{rh}{\Gamma(z,r)} - \frac{r \gamma(\rho,z,r) }{\Gamma(z,r) } \Psi,\\
 z' &= \sqrt{1- \rho^2}, \\ 
 r' &= \rho ,
\end{aligned}\label{finalfull}\right\}
 \end{gather}
where 
\begin{align*}
\gamma(\rho,z,r)=
  \frac{r  \sqrt{1-\rho^2}-z  \rho}{(z^2+r^2)^{3/2}},
 \qquad \Gamma(z,r) = 1 + \frac{z}{\sqrt{r^2+z^2}} ,
\end{align*}
and $\mu  \in C^\infty(\mathbb{R}_+, \mathbb{R}_+)$ satisfies
\begin{align*}
\mu'(\Psi) >0 \; \;  \forall \Psi \in \mathbb{R}_+, \; \;   \lim_{\Psi \rightarrow \infty}\mu(\Psi)= \infty.
\end{align*}
We will consider the phase space given by 
\begin{equation}
M_0= \{ (\rho , h , \Psi, z,r) \in  (-1,1) \times \mathbb{R} \times \mathbb{R} \times \mathbb{R}  \times \mathbb{R}_+     \}.
\end{equation}
We refer to ODE (\ref{finalfull}) as the governing ODE. We denote the vector field corresponding to (\ref{finalfull}) by $V \in C^{\infty}(M, \mathbb{R}^5)$.\\

\subsubsection{Steady tip growth solutions}

Let $x_*:=(\rho_*,h_*,\Psi_*,z_*,r_*)$ be a solution of the governing ODE. Then define $u_*:= \Gamma(z_*,r_*)/(r_* h_*)$. Then we want suitable conditions on $x_*$ such that $r_*,z_*,h_*,\Psi_*,u_*$ satisfy the governing equations (\ref{stressnnnn})-(\ref{zzzz}), (\ref{againage}), the steady tip growth shape and dynamics conditions,  and (\ref{psicondd}). We call a solution $(\rho_*,h_*,\Psi_*,z_*,r_*)$ of the governing ODE (\ref{finalfull}) a steady tip growth solution if it satisfies the following four conditions:
\begin{itemize}
\item[T1] \textbf{Tip limits:} 
\begin{align*}
\lim_{s \rightarrow 0} \rho_*(s) &=1,  &  \lim_{s \rightarrow 0} h_*(s) &=h_0 >0 ,  & \lim_{s \rightarrow 0} \Psi_*(s) &= h_0 z_0^2 ,  \\
  \lim_{s \rightarrow 0}  z_*(s)&= z_0 <0 , &    \lim_{s \rightarrow 0} r_*(s) &=0. 
\end{align*}
\item[T2] \textbf{Analyticity in $r^2$:} There exists a $s_1>0$ and a $G \in C^{\omega}\left( (-a ,a) , \mathbb{R}^4 \right)$ with $a=r_*(s_1)^2$ such that
\begin{align*}
(\rho_*,h_*,\Psi_*,z_*)(s)= G(r_*(s)^2) \qquad \forall s \in (0,s_1).
\end{align*}

\item[T3] \textbf{Global constraints:} For all $s \in (0, \infty)$ the following constraints are satisfied
\begin{align*}
\rho_*'(s) <0, \; \rho_*(s) >0, \; h_*(s)>0. 
\end{align*}
\item[T4] \textbf{Base limits:}  
\begin{align*}
\lim_{s \rightarrow \infty} \rho_*(s) =0,\;   \lim_{s \rightarrow \infty} h_*(s) =h_\infty >0 ,  \;  \lim_{s \rightarrow \infty}  z_*(s)= \infty,  \;  \lim_{s \rightarrow \infty}  r_*(s) =r_\infty>0 .  \\
\end{align*}
\end{itemize}

If $x_*$ is a steady tip growth solution then it follows from Sections \ref{cumff}-\ref{decu} that $r_*,z_*,h_*,\Psi_*,u_*$ satisfy the governing equations (\ref{stressnnnn})-(\ref{zzzz}), (\ref{againage}) and the steady tip growth shape and dynamics conditions.  
Observe that the governing ODE depends on $\mu$ which is  unspecified.

\section{Numerics \label{sec:numerics}}

We are unaware of the existence of biological data on fungal hypha species which can be used to compute $\mu$. Our numerical work suggests that the conditions on the viscosity functions such that governing ODE (\ref{finalfull}) has steady tip growth solutions are not very restrictive. If $\mu$ is a polynomial viscosity function then our numerical results suggest that steady tip growth solutions exist if 
\begin{align}
\mu(\Psi) \in O(\Psi^3), \qquad \mu(0)>0. \label{mu:condiii} 
\end{align}
In addition, for a polynomial viscosity function $\mu$ satisfying (\ref{mu:condiii}) the numerics also suggests that there  exists a smooth one-to-one and onto function $f_\mu: \mathbb{R}_- \rightarrow \mathbb{R}_+$ such that for any $z_0<0$ there exists a steady tip growth solution $(\rho,h,\Psi,z,r)$ with maximal existence interval $(0,\infty)$ such that 
\begin{align*}
 z_0= \lim_{s \rightarrow 0}z(s) , \; \; \;\lim_{s \rightarrow 0}h(s)=f_\mu(z_0)  .
\end{align*}
In terms of the hypha biology this means that the distance from the tip of the cell to the VSC  uniquely determines the cell shape.  \\

\begin{figure}[h]
\begin{center}
\includegraphics[width=15.5cm]{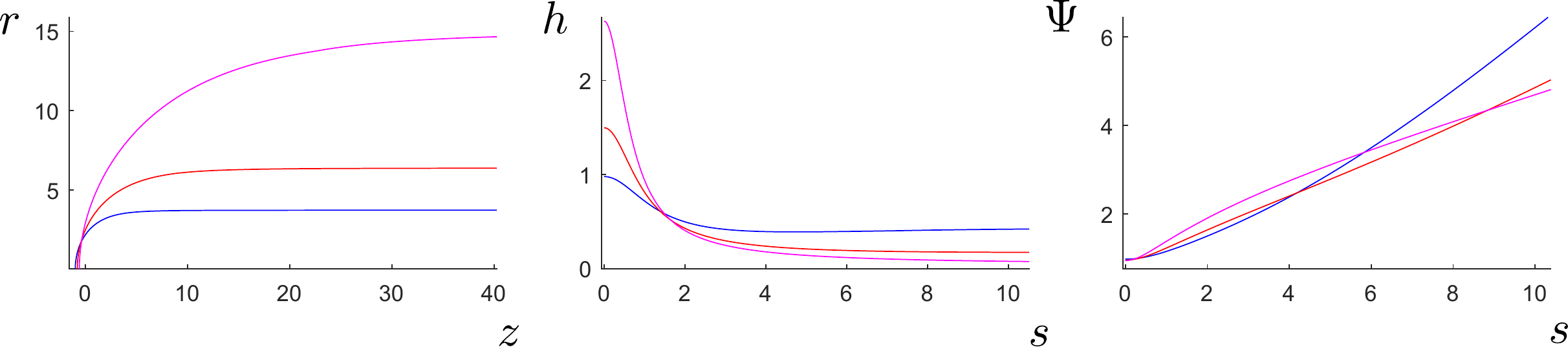}
\caption{ Steady tip growth solutions corresponding to $z_0=-1$, in blue, $z_0=-0.8$, in red, and $z_0=-0.6$, in magenta.  \label{fig:first3} }
\end{center} \end{figure}

Given the viscosity function $\hat{\mu}(\Psi)=1+\Psi^3$ we computed steady tip growth solutions for different $z_0$ in  Figure \ref{fig:first3}. Observe that $\hat{\mu}$ satisfies (\ref{mu:condiii}).

By increasing $z_0$ the tip of the cell becomes more pointed and the width of the cell increases.

\section{Concluding remarks}

We constructed a new model for tip growth of fungal hyphae. Our numerical work suggests that there exist viscosity functions for which the governing ODE (\ref{finalfull}) has steady tip growth solutions. \\

 To  validate the model experimentally the viscosity function needs to be determined. The thickness of the cell wall can be determined \cite{PL03} and the turgor pressure can be measured \cite{MO96}. It is also possible to determine the shape of the cell wall and the location of the Spitzenk\"{o}rper \cite{BG00,BG89}. Hence, we can experimentally determine $h,z,r,u$. Using (\ref{newmassb}) we can determine the VSC rate $C$. Given $h,z,r,C$ we can compute $\Psi$ using the integral equation (\ref{againage}). Finally, we can use the $h$-equation  or the $\rho$-equation in (\ref{finalfull}) to  determine the viscosity function $\mu$. Given $\mu$ we can then proceed to verify if the governing ODE (\ref{finalfull}) has a steady tip growth solution which approximates the experimentally determined cell growth.  From a biological perspective the determination of these variables would be a suitable topic for future work. In case this model does not match the experimental results the viscosity of the cell wall will not have a dominant role in tip growth of fungal hyphae. Studies on the material properties of fungal cell walls are extremely useful to obtain accurate models. It might also be useful to study the fungal cell wall at a molecular level as the chemical processes involving the cell wall polymers determine its material structure. This can be studied in an implicit way by observing how gene mutations affect hypha morphogenesis. Pioneering work in this field has been done by Gordon et al. \cite{GOA00,GAK00}

 Giving an existence proof for steady tip growth solutions would be interesting from a theoretical point of view but turns out to be complicated as (\ref{finalfull}) is a 5 dimensional non-linear ODE without obvious structural properties to be exploited.

Furthermore, steady tip growth solutions correspond to travelling wave profiles of an underlying PDE to the governing ODE (\ref{finalfull}). Should travelling wave profiles exist, in the sense of steady tip growth solutions,  it is a natural question whether these are stable with respect to small perturbations. Answering this question, however, would demand methods of PDE rather than ODE analysis.

\subsection*{Acknowledgement} 
This research was partly funded by a PhD grant of the NWO Cluster ``Nonlinear Dynamics in  Natural Systems''(NDNS+) and by NWO VICI grant 639.033.008.

\bibliographystyle{plain}
\bibliography{ref}

\begin{thebibliography}{10}

\bibitem{BG00}
S.~Bartnicki-Garcia, C.~Bracker, G.~Gierz, R.~Lopez-Franco, and H.S. Lu.
\newblock Mapping the growth of fungal hyphae: Orthogonal cell wall expansion
  during tip growth and the role of turgor pressure.
\newblock {\em Biophysics Journal}, 79(5):2382--2390, 2000.

\bibitem{BG89}
S.~Bartnicki-Garcia, F.~Hergert, and G.~Gierz.
\newblock Computer simulation of fungal morphogenesis and the mathematical
  basis for hyphal (tip) growth.
\newblock {\em Protoplasma}, 153(1-2):46--57, 1989.

\bibitem{CM09}
O.~Campas and L.~Mahadevan.
\newblock Shape and dynamics of tip growing cells.
\newblock {\em Current Biology}, 19:2102--2107, 2009.

\bibitem{DU04}
J.~Dumais, S.R. Long, and S.L. Shaw.
\newblock The mechanics of surface expansion anisotropy in medicago tuncutala
  root hairs.
\newblock {\em Plant Phys.}, 136:3266--3275, 2004.

\bibitem{EGG11}
E.~Eggen, M.N.~de Keijzer, and B.M. Mulder.
\newblock Self-regulation in tip growth: The role of cell wall ageing.
\newblock {\em J. of Theo. Bio.}, 283(1):113--121, 2011.

\bibitem{GI55}
M.~Girbardt.
\newblock Lebendbeobachtungen on {P}olystictus versicolor ({L}).
\newblock {\em Flora}, 142:540--563, 1955.

\bibitem{GI57}
M.~Girbardt.
\newblock Der {S}pitzenk\"{o}rper von {P}olystictus versicolor ({L}).
\newblock {\em Planta}, 50:47--59, 1957.

\bibitem{GI69}
M.~Girbardt.
\newblock Die {U}ltrastruktur der {A}pikalregion von {P}ilzhyphen.
\newblock {\em Proplasma}, 67:413--441, 1969.

\bibitem{GOA00}
C.L. Gordon, D.B. Archer, D.J. Jeenes, J.H. Doonan, B.~Wells, A.P.J. Trinci,
  and G.D. Robson.
\newblock A glucoamylase:gfp gene fusion to study protein secretion by
  individual hyphae of aspergillus niger.
\newblock {\em J Microbiol Methods}, 42:39--48, 2000.

\bibitem{GAK00}
C.L. Gordon, V.~Khalaj, A.F.J. Ram, D.B. Archer, J.L. Brookman, A.P.J. Trinci,
  D.~Jeenes, J.H. Doonan, B.~Wells, P.J. Punt, C.~van~den Hondel, and G.D.
  Robson.
\newblock Glucoamylase: green fluorescent protein fusions to monitor protein
  secretion in aspergillus niger.
\newblock {\em Microbiology UK}, 146:415--426, 2000.

\bibitem{GO17}
A.~Goriely.
\newblock {\em The mathematics and mechanics of biological growth}.
\newblock Springer, New York, 2017.

\bibitem{GO10}
A.~Goriely, M.~Tabor, and A.~Tongen.
\newblock A morpho-elastic model of hyphal tip growth in filamentous organisms.
\newblock In K.~Garikipati and E.~Arrudar, editors, {\em IUTAM Symposium on
  Cellular, Molecular and Tissue Mechanics}, pages 232--260. Springer,
  Dordrecht, 2004.

\bibitem{HA02}
F.M. Harold.
\newblock Force and compliance: rethinking morphogenesis in walled cells.
\newblock {\em Fungal Genetics and Biology}, 37:271--282, 2002.

\bibitem{HW94}
P.D. Howell.
\newblock {\em Extensional thin layer flows}.
\newblock PhD thesis, University of Oxford, 1994.

\bibitem{HUD92}
H.J. Hudson.
\newblock {\em Fungal biology}.
\newblock Cambridge University Press, Cambridge, 1992.

\bibitem{JONGCOMFOS}
T.~de Jong, G.~Prokert, and J.~Hulshof.
\newblock A new model for fungal hyphae growth using the thin viscous sheet
  equations.
\newblock In {\em Proceedings of the international conference Comfos}, 2016.

\bibitem{KEIJ09}
M.N.~de Keijzer, A.M. Emons, and B.M. Mulder.
\newblock Modelling tip growth: pushing ahead.
\newblock In T.~Ketelaar A.M.~Emons, editor, {\em Root Hairs}, pages 103--122.
  Springer, Berlin, 2009.

\bibitem{KOCH94}
A.~Koch.
\newblock The problem of hyphal growth in streptomycetes and fungi.
\newblock {\em J.Theor.Biol.}, 171:137--150, 1994.

\bibitem{LEW07}
R.R. Lew and N.N. Levina.
\newblock Turgor regulation in osmosensitive cut mutant of neurospora crassa.
\newblock {\em Microbiology}, 151:1530--1537, 2007.

\bibitem{LEW04}
R.R. Lew, N.N. Levina, S.K. Walker, and A.~Garrill.
\newblock Turgor regulation in hyphal organisms.
\newblock {\em Fungal Genetics and Biology}, 15:1007--1015, 2005.

\bibitem{MCC68}
K.~McClure, D.~Park, and P.M. Robinson.
\newblock Apical organization in the somatic hyphae of fungi.
\newblock {\em J. Gen. Microbiol.}, 50:177--182, 1986.

\bibitem{MCD00}
D.P. McDaniel and R.W. Roberson.
\newblock Microtubules are required for motility and positioning of vesicles
  and mitochondria in hyphal tip cells of allomyces macrogynus.
\newblock {\em Fungal Genetics and Biology}, 31:233--244, 2000.

\bibitem{MO96}
N.P. Money.
\newblock Confirmation of a link between fungal pigmentation, turgor pressure,
  and pathogenicity using a new method of turgor pressurement.
\newblock {\em Fungal Genetics and Biology}, 2:221--227, 1996.

\bibitem{MO08}
N.P. Money.
\newblock Insights on the mechanics of hyphal growth.
\newblock {\em Fungal Biology Rev.}, 22:71--76, 2008.

\bibitem{MOHA93}
N.P. Money and F.M. Harold.
\newblock Two water moulds can grow without measurable turgor pressure.
\newblock {\em Planta}, 190:426--430, 1993.

\bibitem{PL03}
B.C. Paul, H.~Ma, L.A. Snook, and T.E.S. Dahms.
\newblock High resolution imaging and force spectroscopy of fungal hyphal cells
  by atomic force microscopy.
\newblock In V.K~Gupta et. al., editor, {\em Laboratory protocols in fungal
  biology}, pages 151--160. Springer, New York, 2013.

\bibitem{PIP68}
A.C. Pipkin.
\newblock Integration of an equation in membrance theory.
\newblock {\em ZAMP}, 19(5):818--819, 1968.

\bibitem{RI14}
M.~Riquelme, E.L. Bredeweg, O.~Callejas-Negrete, R.W. Roberson, S.~Ludwig,
  A.~Beltr\'{a}n-Aquilar, S.~Seiler, P.~Novick, and M.~Freitag.
\newblock The neurospora crassa exocyst complex tethers spitzenk\"{o}rper
  vesicles to the apical plasma membrane during polarized growth.
\newblock {\em Mol. Biol. Cell.}, 25(8):1312--1326, 2014.

\bibitem{STE07}
G.~Steinberg.
\newblock Hyphal growth: A story of motors, lipids, and the spitzenk\"{o}rper.
\newblock {\em Eukaryotic Cell}, 6(3):351--360, 2007.

\bibitem{WES83}
J.G. Wessels, J.H. Sietsma, and A.S. Sonnenberg.
\newblock Wall synthesis and assembly during hyphal morphogenesis in
  schizophyllum commune.
\newblock {\em J. of Gen. Microbio.}, 129:1607--1616, 1983.

\end{thebibliography}

\end{document}